# Mid-infrared LEDs based on lattice-mismatched hybrid IV-VI/III-V heterojunctions

Jarod E. Meyer[1], Biridiana Rodriguez[2], Leland Nordin[2,3*], and Kunal Mukherjee[1†]

[1]Department of Materials Science and Engineering, Stanford University, Stanford, CA 94305, USA

[2]CREOL, The College of Optics and Photonics, University of Central Florida, Orlando, FL 32816, USA

[3]Department of Materials Science and Engineering, University of Central Florida, Orlando, FL 32816, USA

## ABSTRACT

Light-emitting diodes (LEDs) can bridge the gap between narrow linewidth, expensive lasers and broadband, inefficient thermal globars for low-cost chemical sensing in the mid-infrared (mid-IR). However, the efficiency of III-V based mid-IR LEDs at room temperature is low, primarily limited by strong nonradiative Auger-Meitner recombination that is only partially overcome with complex quantum-engineered active regions. Here, we exploit the intrinsically low Auger-Meitner recombination rates of the IV-VI semiconductors PbSe and PbSnSe, while leveraging the mature III–V platform through the fabrication of hybrid heterojunctions that mediate the ~8% lattice mismatch to GaAs. Electrically injected n-PbSe/p-GaAs LEDs emit at 3.8 µm with output powers up to 400 µW under pulsed operation and a peak wall plug efficiency of 0.08% at room temperature, approaching the performance of commercial III-V LEDs at similar wavelengths. Incorporating 7% Sn extends the emission to 5 µm in GeSe/PbSnSe/GaAs LEDs with output powers up to 45 µW. Notably, both devices operate despite threading dislocation densities on the order of $10^9$ cm$^{-2}$, underscoring the potential of hybrid IV-VI/III-V heterojunction architectures. We show that combining the complementary advantages of IV-VI and III-V semiconductors offers a simple and efficient mid-IR optoelectronic platform for a rapidly expanding set of applications.

* leland.nordin@ucf.edu

† kunalm@stanford.edu

## 1. INTRODUCTION

PbSe is a semiconductor in the rocksalt-structured IV-VI family that is important for mid-infrared (mid-IR, 3–30 μm) optoelectronic devices, and merits further attention especially for light emitters. It has an innately low nonradiative Auger-Meitner recombination coefficient in comparison to other bulk III-V and II-VI narrow gap semiconductors,[1] which is the leading loss mechanism for LEDs and lasers at these long wavelengths and high operating temperature detectors.[2] It also has unconventional bonding leading to, among other features, a high static dielectric constant that may impart a degree of defect tolerance.[3] Furthermore, by alloying Sn to form PbSnSe, the emission wavelength can be tuned across the entire mid-IR spectrum.[4] Taken together, these features potentially enable high quantum efficiency emitters and detectors based on IV-VI semiconductors to be monolithically integrated even on lattice and crystal-structure mismatched substrates like Si, Ge, and GaAs.[1] In fact, polycrystalline PbSe (and PbS) detectors, (when appropriately sensitized), yield some of the best cost to performance capabilities for uncooled mid-IR detection as a result of these attractive material qualities.[5]

Despite the technological and commercial relevance of PbSe on the detection side, little recent work has focused on advancing PbSe based mid-IR lasers and LEDs. With emerging applications in thermography, environmental emissions monitoring, gas-leak detection, and industrial process control requiring widespread deployment of mid-IR optical gas sensors, there is great interest in the development of efficient, simple, low-cost, uncooled light sources,[6] qualities which incumbent technologies do not offer. On the one hand, III-V semiconductor light emitters offer high performance, but require thick epitaxial active regions and/or highly engineered quantum cascade or interband cascade type-II superlattice structures to reduce the Auger-Meitner coefficient, leading to higher costs and fabrication complexity.[7–11] Thermal blackbody emitters, on the other hand, offer simple fabrication but suffer from poor emission efficiency, limited by their fundamentally equilibrium output, and concomitantly low output powers, especially at longer wavelengths.[6] A simpler mid-IR LED could fill the gap between expensive III-V lasers and thermal sources, and this grand challenge has prompted extensive work into exploring LEDs based not only on III-V semiconductors but also newer materials like black phosphorus and intraband HgTe/CdSe colloidal quantum dots.[12–15] In this context, type-I band-aligned LEDs based on PbSe and its alloys are worth a closer look.

Historically, PbSe-(and PbTe-)based lasers were widely used for mid-IR emission before the invention of III-V cascade lasers,[6] yet continuous wave lasing from electrically pumped IV-VI devices at room temperature remains a challenge.[16] The precise sources of loss are not established. Poor thermal conductivity of native IV-VI (4.2 W/m–K) and $BaF_2/CaF_2$ substrates (20 W/m-K), leading to severe self-heating effects, and the poor optoelectronic quality of wide-gap barrier layers such as PbSrSe and PbEuSe used for carrier confinement are factors that may limit the performance of IV-VI lasers.[1,17,18] Unlike PbSe lasers, only a few reports explore the spontaneous emission behavior of optically pumped and electrically pumped PbSe mid-IR LEDs.[17,19] The reported output power of electrically injected 4.6 µm PbSSe LEDs are only a few hundred nW at room temperature.[20] For bulk PbSnSe LEDs, only a few nW of output power was reported even with operation at cryogenic temperatures.[21] These results raise the central question: can the poor performance of IV-VI LEDs in prior work be overcome through materials engineering and integration, or was it fundamentally tied to the IV-VI material system despite its attractive bulk electronic properties?

Addressing this question has motivated recent efforts to improve IV-VI material performance through integration, particularly a IV-VI/III-V platform via cube-on-cube heteroepitaxial growth.[22–26] For light emitters especially, the superior thermal conductivity and mechanical/chemical stability of III-V substrates are attractive.[24] Although, the large lattice and crystal structure mismatch introduces crystal defects, bright 4 µm photoluminescence (PL) with estimated internal quantum efficiencies (IQE) of about 5% at room temperature has been observed in such heteroepitaxial PbSe thin films where the threading dislocation density (TDD) is in the $10^9$ $cm^{-2}$ range.[27,28] This value of IQE is already close to that achieved by engineered III-V heterostructures at room temperature at similar wavelengths.[9,29] When combined with a naturally low temperature for IV-VI epitaxy below 350 °C, the easy integration of high optoelectronic quality PbSe on III-V substrates opens doors for leveraging substrate band offsets and doping as active portions of IV-VI/III-V heterojunction, optoelectronic devices.[30] Indeed, PbSe heterojunction diodes were recently demonstrated using the related group-IV Ge substrate, where the type II band alignment was utilized in n-PbSe/p-Ge photovoltaic detectors.[31] The possibility of using PbSnSe for wavelength extension deeper into the mid-IR was also recently demonstrated in films grown on GaAs substrates. PbSe/PbSnSe thin film heterostructures grown on GaAs exhibited bright PL out to 8 μm at room temperature after in-situ surface passivation with GeSe, to block defects related to oxidation.[28]

In this work, we demonstrate room temperature operating mid-IR LEDs based on n-i-p PbSe/GaAs heterostructures, leveraging the type-I band alignment between PbSe and the wide-gap GaAs substrate. The heterostructure diodes show strong rectification behavior, and 4 μm EL with output powers up to 400 μW and peak wall plug efficiencies of 0.08% under 1% duty cycle operation, measured at room temperature. GeSe/PbSnSe/GaAs LEDs were also fabricated to tune the emission wavelength deeper into the mid-IR. While the insulating GeSe layer leads to higher series resistance and device heating effects, EL at 5 μm with output powers up to 45 μW were measured for a 7% Sn PbSnSe LED, demonstrating the feasibility of tuning emission wavelength by alloy composition. Overall, these results highlight the strong potential of electrically injected IV-VI LEDs, integrated on mature III-V GaAs substrates, for applications including, but not limited to, low-cost mid-IR gas sensing applications.

## 2. METHODS

*Epitaxial Growth*

Device epitaxial layers were grown in a Riber Compact 21 Molecular beam epitaxy (MBE) system utilizing 6N purity PbSe, SnSe, and GeSe compound source dual zone effusion cells, and 6N pure elemental Se and Sb valved cracker sources. An optical pyrometer calibrated to the GaAs native oxide desorption at 600 °C was used to measure temperature, and Reflection high energy electron diffraction (RHEED) with a 15 kV electron gun was used to monitor the growth. For PbSe and PbSnSe growth, a PbSe beam-equivalent-pressure (BEP) of $3 \times 10^{-7}$ torr was utilized, corresponding to a growth rate of 0.42 Å/s for (001)-oriented epitaxial layers in the unity sticking regime. Ternary PbSnSe layers were grown by supplying additional flux from the SnSe cell, with the alloy composition set by tuning the PbSe:SnSe flux ratio. Epi-ready p-type GaAs (001) ¼ 2" and 2" substrates were mounted indium-free in Mo platens and degassed in a buffer chamber prior to loading into the MBE growth chamber. The GaAs native oxide was desorbed at 600 °C under a Se BEP of $2 \times 10^{-7}$ torr until a Se-terminated GaAs RHEED pattern was observed.[32] Following native oxide removal, the GaAs surface was exposed to a PbSe flux at 420 °C for 30 s to promote single-orientation (001) nucleation.[24]

For the PbSe LED, a 25 nm PbSe nucleation layer was deposited at 315 °C. RHEED was initially spotty but transitioned to a streaky 1 × 1 pattern by the end of the nucleation layer step, (Figure S1). Then, the sample was cooled to ~ 170 °C and 450 nm of unintentionally doped PbSe was grown, followed by 100 nm of Sb-doped, n-type PbSe, with Sb supplied from a valved cracker cell with a BEP of $< 10^{-9}$ torr.[33] Sb-doped PbSe thin films grown on semi-insulating GaAs substrates showed n-type doping concentrations of $10^{19}$ $cm^{-3}$, and hall mobilities in the 300 – 500 $cm^2$/V-s range, while unintentionally doped bare PbSe films showed p-type doping in the few $10^{18}$ $cm^{-3}$ range, stemming from exposure to atmospheric oxygen after removal from the MBE chamber (Figure S3).[34] The growth remained single orientation at 170°C, but some roughening was observed in RHEED, as the streaky 1 × 1 pattern became slightly spottier over time.

For the PbSnSe LED, a 25 nm nucleation layer of PbSe was deposited at 330 °C until a streaky 1 × 1 RHEED pattern was observed, (Figure S1), followed by cooling down to 190 °C. Then, an additional 37.5 nm of PbSe was deposited, followed by 150 nm of 7% Sn PbSnSe. The last 10 nm of PbSnSe was doped with Sb. To prevent oxidation of the PbSnSe layer, which was previously found to severely degrade the optoelectronic properties, a 60 nm amorphous GeSe cap was deposited in-situ.[28] Only the last 10 nm of the PbSnSe layer was doped with Sb to form an $n^+$ region, as previously unintentionally doped PbSe and PbSnSe layers capped in-situ with amorphous GeSe were found to be n-type in the $10^{18}$ $cm^{-3}$ range, (due to metal-rich growth conditions), as oxidation was prevented from switching the carrier type to p-type.[28] Similar to the PbSe LED epitaxial stack, some roughening was observed in RHEED during the low temperature PbSe/PbSnSe deposition, but the growth remained single orientation. The GeSe capping layer was grown using a GeSe BEP of $3 \times 10^{-7}$ torr, corresponding to an amorphous film growth rate of ~ 0.33 Å/s, at temperatures < 160 °C, after switching off the substrate heater power. RHEED transitioned to a hazy, amorphous glow after deposition of the GeSe.

*Post-growth processing and characterization*

After growth, samples were cleaved into 1 cm × 1 cm pieces, cleaned with acetone, isopropyl alcohol, and then dried with an $N_2$ gun. Pieces were loaded into a PlasmaTherm Versaline high density plasma chemical vapor deposition system, where 110 – 120 nm of $SiO_2$ was deposited at a substrate temperature of 90 °C. $SiO_2$ thickness was confirmed via ellipsometry by simultaneously depositing $SiO_2$ on pieces of prime-grade Silicon. After the dielectric film deposition, samples were rapid thermal annealed (RTA) in an Allwin 610 RTA system at 375 °C for 300 s under $N_2$. During the RTA, the $SiO_2$ capped PbSe/PbSnSe epilayers were sandwiched between the polished sides of two pieces of Silicon. In previous work, RTA under such conditions was found to significantly improve luminescence efficiency for films grown at low temperatures.[35] For the PbSnSe LED, the RTA process also led to crystallization of the amorphous GeSe cap layer; peaks corresponding to orthorhombic GeSe (c-GeSe) were observed in open detector, symmetric X-ray diffraction (XRD) scans after annealing.[28]

High resolution XRD scans were taken using a Panalytical X'Pert PRO MRD system with Cu Ka1 radiation. $2\theta$ - $\omega$ scans were taken in triple-axis configuration. Symmetric (004) and skew-symmetric (224) rocking curves, ($\omega$ scans), were measured in the double-axis, "open-detector" configuration. Room temperature Hall measurements were conducted in the Van der Pauw configuration with a magnetic field of 0.9 T. Calibration samples consisting of 80 nm PbSe thin films, with and without Sb doping and with and without 60 nm GeSe capping layers, were grown on semi-insulating GaAs (001) substrates. The samples were cleaved into $5 \times 5$ mm squares and metallized with Indium soldered contacts on all 4 corners. Quasi-continuous-wave photoluminescence (PL) measurements were conducted using a 10 kHz electrically modulated 808 nm laser, a $LN_2$ cooled HgCdTe detector, and a Bruker Invenio Fourier-transform infrared (FTIR) spectrometer operating in step-scan mode. Further details of the setup have been described previously.[28,35]

*LED fabrication*

LEDs were fabricated using standard optical photolithography, a combination of dry etching and citric acid-based wet etching, and electron beam evaporation for metal contact deposition. For the top contact, the $SiO_2$ capping layer was first dry etched, followed by deposition of 15/200 nm Ti/Au as a contact to n-PbSe/GeSe for the PbSe and PbSnSe LEDs, respectively. Mesas were defined using Ar-dry etching down to the p-GaAs substrate, followed by wet etching 300 nm of GaAs in a 5:1:1 $C_6H_8O_7$:$H_2O_2$:$H_2O$ citric acid mixture.[36] A 75/75/200 nm Ti/Pd/Au contact metal stack was then evaporated for the p-GaAs contact.[37] Without citric acid wet etching, as-deposited Ti/Pd/Au contacts were found to be Schottky-type to the p-GaAs substrate, likely due to Ar-dry etching induced surface damage. After wet etching through the damaged GaAs layer, however, as-deposited Ti/Pd/Au contacts were found to be ohmic. The metal contacts were not annealed after deposition.

*Device testing*

Current-Voltage (I-V) measurements on LEDs were conducted using a Keithley 2400 source measurement unit. 4-wire measurements were used to remove the influence of probe-resistance from the I-V data. I-V characteristics of the n-PbSe, p-GaAs, and GeSe metal contacts were measured by the Transmission line method (TLM). Room temperature EL spectra were measured in the same PL setup but using an ILX Lightwave LDC-3744B laser diode controller to apply 5 kHz, 50% duty cycle square wave current pulses to devices under test. Quantitative light-current (L-I) measurements of LED output power were conducted at 1 kHz, 1% duty cycle in a 1:1 imaging setup using an Agilent 8114A pulse generator, a HgCdTe detector, and a commercial Hamamatsu L15895-0430MA LED as a calibration source.[38] The qualitative L-I characteristics of the commercial LED was first measured, and then the commercial LED datasheet was used to convert LED L-I to upper hemisphere output power. The slightly non-Lambertian emission profile of the commercial LED was not considered, but the wavelength dependent spectral response of the detector was corrected for. The detector spectral response was measured using a FTIR spectrometer. Light from the FTIR's internal globar source was focused onto the detector using a 1.5-inch diameter, 4-inch focal length gold-coated parabolic mirror. The measured spectral response was normalized to that of a spectrally flat pyroelectric detector placed in the same optical setup, effectively removing the spectral weighting associated with the FTIR beamsplitter and internal blackbody source. To account for the frequency dependence of the pyroelectric detector, a transfer function was fitted based on a series of detector velocity measurements.[39]

## 3. RESULTS

### 3.1. Device structures and material characterization

Figure 1a) shows the schematic PbSe and PbSnSe hybrid LED structures explored in this work. These are nominally p-i-n junctions typical for LEDs but uniquely harness the GaAs template not just as a substrate for growth but, perhaps more importantly, also as a part of the electrical junction. We modify the heterostructure with a GeSe capping/diffusion-barrier layer for the longer wavelength PbSnSe device to circumvent the tendency of Sn-related oxidation that quenches luminescence. No oxygen-related diffusion barriers were found necessary for PbSe. It is important that the heterojunction band alignments between PbSe/GaAs and PbSnSe/GeSe be type-I in nature for minority carrier injection. Given the thin active layers in this work, a type-II or type-III alignment could lead to rapid separation of electron-hole pairs, quenching luminescence. Figure 1b) illustrates tentative band alignment using literature reported values of the electron affinities and bandgaps for GaAs, PbSe, and GeSe, suggesting that the wide-gap GaAs and GeSe layers are likely to form type-I barrier layers for PbSe.[31,40] Type-I alignment between GaAs/GeSe and 7% Sn PbSnSe is also expected, as the relatively dilute Sn alloying should not shift the position of the bands much with respect to the two much wider gap semiconductors. Such layers, therefore, should be beneficial in providing electron and hole confinement during LED operation, provided they do not introduce large series resistance losses. Prior PL measurements on sub 100 nm films of PbSe and PbSnSe on GaAs, InAs, and GaSb tentatively suggest favorable alignment in practice as well. We observed bright PL from PbSe and GeSe-capped PbSnSe on 8%-lattice mismatched GaAs but dim PL from PbSe on 1.5%-mismatched InAs and 0.7%-mismatched GaSb, likely due to photocarrier flow into the narrower bandgap III-V substrates.[27,28] Further work is needed to independently confirm this assignment, as heterovalent interfaces and local chemistry are known to modify the band alignment.[41] It is also not known if the heterojunction alignment between PbSe and dilute PbSnSe is type-I or type-II, though heterojunctions in the related PbTe/PbSnTe material system have been found to possess type-II alignment.[42]

We mediate the 8%-lattice mismatch and achieve (001) cube-on-cube epitaxy of the IV-VI layers on GaAs(001) using a nucleation strategy that involves a high temperature exposure to PbSe flux followed by a lower temperature buffer.[24] The nucleation sequence, first developed for bare PbSe films, is critical as the heterointerface is part of the electrical junction. Special considerations are additionally needed for thicker layers necessary in LEDs as IV-VI and III-V semiconductors have a large thermal expansion coefficient mismatch (~13 ppm/K),[22,27] which leads to cracking in epilayers above 200 nm in (001)-orientations when grown at 300 °C.[25] We overcome this limit by conducting most of the epitaxial growth at 190 °C, followed by a short 300 s thermal anneal at 375 °C after ex-situ capping with chemical vapor deposited $SiO_2$ at 90 °C. The $SiO_2$ capped RTA procedure allows films thicker than 400 nm to be synthesized without significant cracking and recovers PL intensity comparable to conventional 300 °C grown samples. Figure 1c) plots the triple-axis, symmetric 2θ/ω scans of the two device epilayers after RTA near the PbSe/GaAs (004) reflections. The peak of the PbSnSe LED epilayer is shifted with respect to the PbSe (004) due to the reduction in lattice constant with Sn alloying.[43] The full-width-at-half-maximum (FWHM) of both symmetric (004) and skew-symmetric (224) XRD rocking curves is seen to decrease with RTA (Figure S2). TDD is calculated from the FWHM, β, of skew-symmetric (224) rocking curves before and after RTA via[35,44,45]:

$$TDD = \frac{\beta_{224}^2}{4.35b^2}. \tag{1}$$

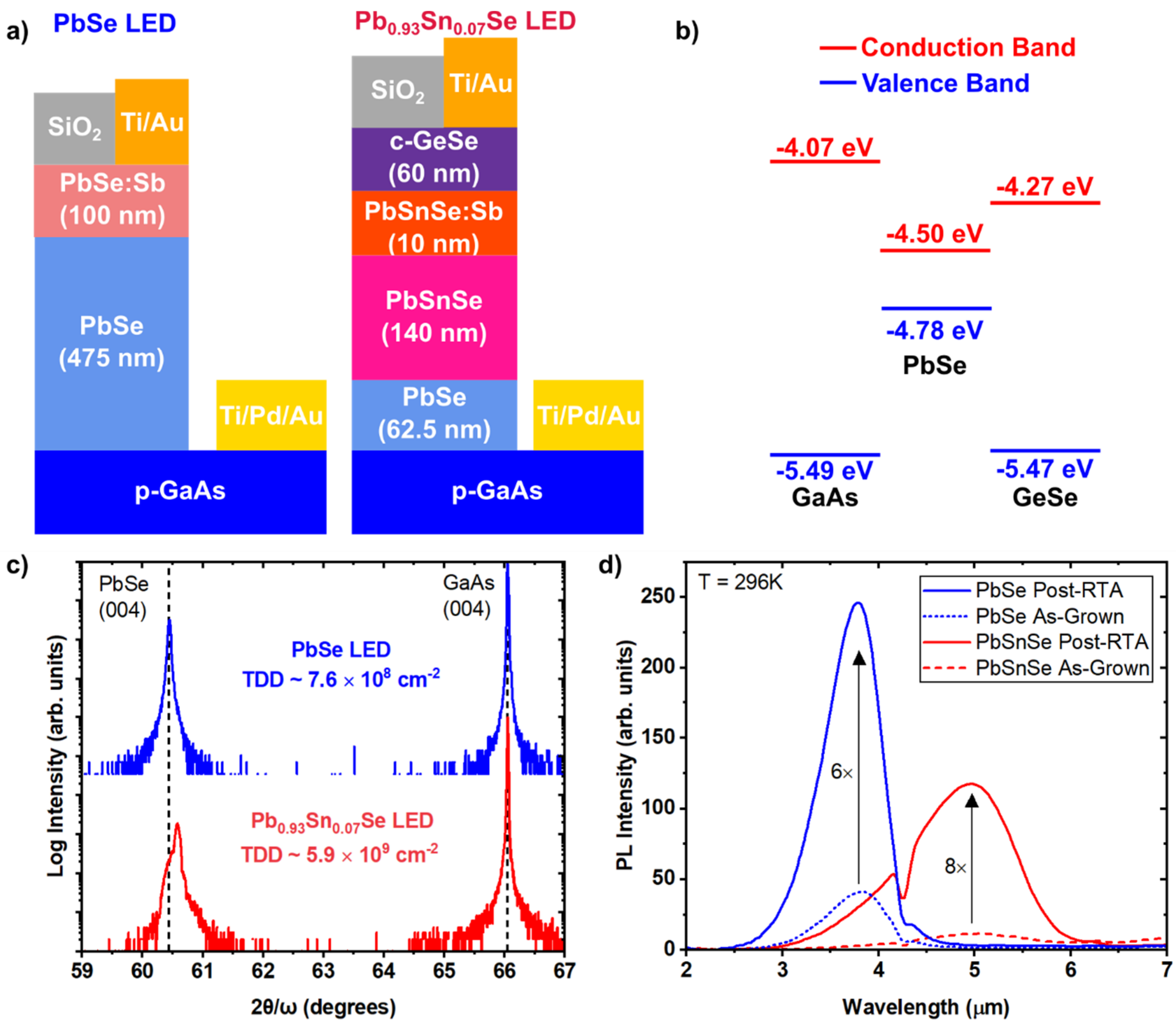

**Figure 1. a)** Schematic cross sections showing the layer structures for the PbSe and PbSnSe LEDs studied in this work. **b)** Electron affinity aligned band gaps for GaAs, PbSe, and GeSe taken from literature.[31,40] **c)** Triple-axis, symmetric 2θ/ω near the PbSe and GaAs (004) reflections for the PbSe and PbSnSe LEDs after $SiO_2$ capping and rapid thermal annealing. Threading dislocation densities (TDD), determined from skew symmetric, open detector (224) rocking curves, are also stated. **d)** High-injection, room temperature photoluminescence (PL) spectra of PbSe and PbSnSe LED epilayers before and after annealing. PL intensity in both samples is greatly improved after annealing, by 6× and 8× respectively, with no significant shift in emission wavelength.

Here, $b$ is the Burgers vector magnitude for PbSe and is equal to $\frac{a}{2}[110]$. For the PbSe sample, TDD was seen to decrease from $2.9 \times 10^9$ to $7.6 \times 10^8$ $cm^{-2}$ with RTA. For the PbSnSe sample, a much higher TDD of $8.1 \times 10^9$ $cm^{-2}$ was derived as-grown, decreasing to $5.9 \times 10^9$ $cm^{-2}$ after RTA. The PbSnSe layers exceed the critical thickness for dislocation nucleation with respect to the PbSe buffer and may result in the observed increase in TDD.

Room temperature, high-injection PL spectra for the PbSe and PbSnSe epilayers before and after RTA are plotted in Figure 1d). Strong PL is recovered after RTA at 3.8 and 5 μm for the PbSe and PbSnSe samples, respectively. No shift in wavelength with annealing is observed for the PbSnSe sample, suggesting the annealing temperature of 375 °C is low enough to prevent significant interdiffusion of the PbSe/PbSnSe/GeSe layers. PL intensity after annealing is improved by 6× and 8× for the PbSe and PbSnSe samples respectively, though it is not correlated with changes in the TDD. While PL increases by nearly an order of magnitude for the PbSnSe sample, the TDD only decreases by about 30%. Furthermore, despite a roughly 8× higher TDD than the PbSe sample after RTA, the PbSnSe sample is only dimmer by about a factor of 2×. This aligns with previous work suggesting that nonradiative Shockley-Read-Hall (SRH) recombination at point defects is dominant in IV-VI semiconductors, rather than specifically at dislocations.[35] For the PbSnSe sample, the crystallization of the GeSe capping layer with RTA may also lead to the increase in PL intensity by reducing the density of nonradiative interface recombination states.

### 3.2. PbSe LED characterization

The epilayers after RTA were fabricated into devices. Figure 2a) shows an optical microscopy image of a set of fabricated PbSe LEDs with varying metal contact coverages. TLM measurements on the Ti/Au and Ti/Pd/Au contacts, to n-PbSe and p-GaAs respectively, were ohmic (Figure S4a-b). Figure 2b) plots the current density-voltage (J-V) curves for fully covered PbSe LED mesas from -0.5 to 0.7 V. We observe strong rectification greater than 100× between +/- 0.5 V, a turn-on behavior close to the bandgap of PbSe, and leakage currents were below 1A/$cm^2$ at -0.5 V for all mesa sizes. Importantly, we find this leakage current at -0.5 V is about a factor of 5× lower than for recent polycrystalline PbSe/CdSe photovoltaic detectors at room temperature, and also epitaxial AlInSb Mid-IR LEDs.[46,47] The series resistance for a 200 × 350 μm diode was found to be roughly 0.2 Ω. After correcting for series and shunt resistances, the ideality factor was derived by fitting the I-V data from 0 – 0.5 V to the ideal diode equation, $I = I_0(e^{\frac{qV}{nkT}} - 1)$, where n is the ideality factor.[48] A value of 1.66 was extracted for n, (close to n = 2), indicating significant SRH recombination in the space charge region.[49]

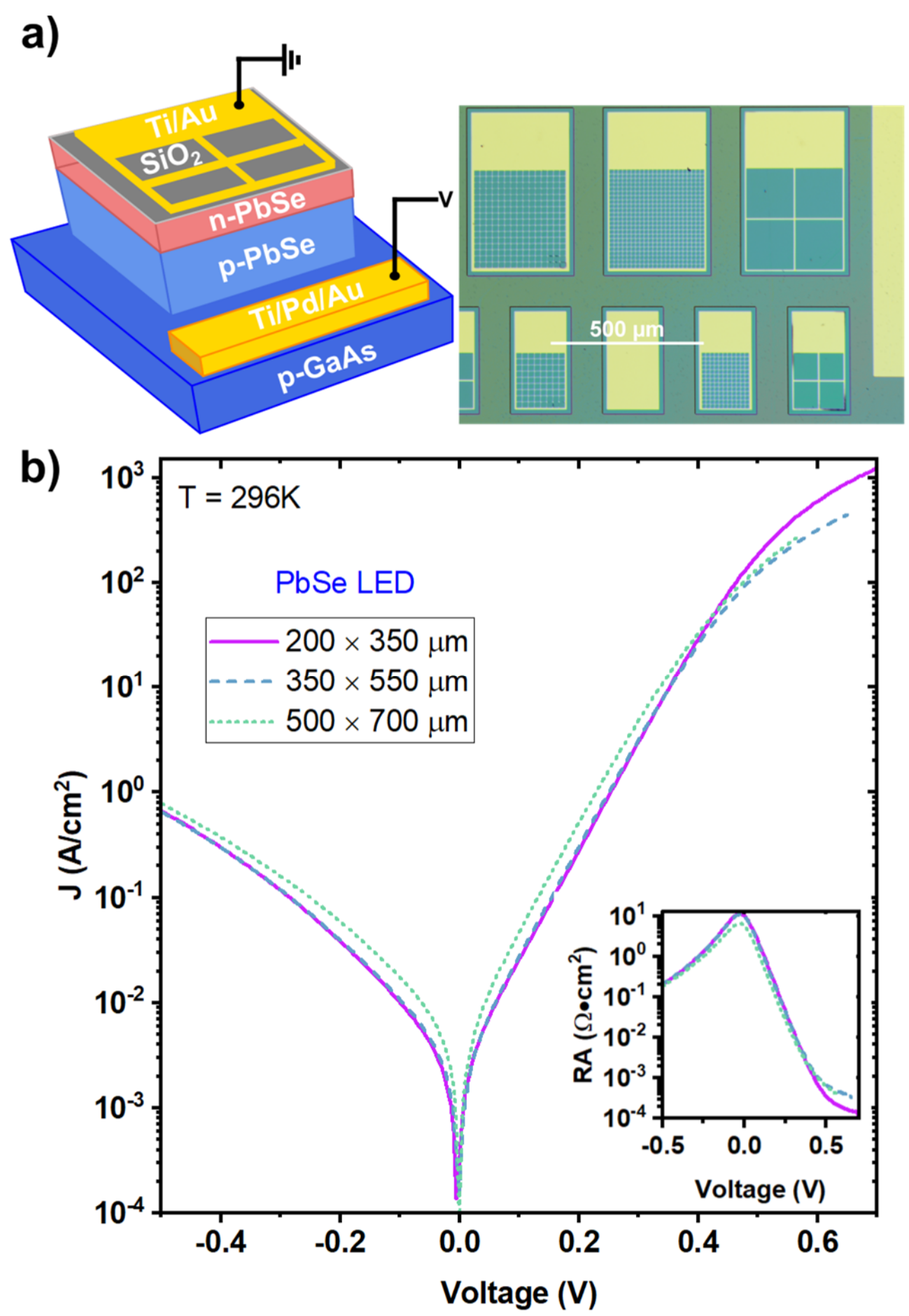

**Figure 2. a)** Device schematic alongside optical microscopy image of fabricated PbSe/GaAs diodes with varying top metal contact styles. **b)** Room temperature J-V curves for PbSe diodes of 3 different mesa sizes, with the applied voltage swept from -0.5 – 0.7 V. The leakage current at -0.5 V is seen to be less than $1 A/cm^2$ for the PbSe/GaAs diodes, with a rectification ratio of roughly 100× between +/- 0.5 V. The inset shows the differential resistance area products for mesas fully covered by metal.

The inset to Figure 2b) plots the differential resistance area products for three PbSe LED mesa sizes at room temperature. A resistance area product at zero bias, $R_0A$, of 10.35 Ω $cm^2$ is derived for the smallest size PbSe LED at room temperature, decreasing to 10.00 and 5.99 Ω $cm^2$ for the medium and large size mesas, respectively. The $R_0A$ for the diodes is much larger than theoretical and experimental values for bulk and MBE-grown homojunction PbSe diodes,[50,51] likely due to a heterojunction effect with the wide-gap GaAs layer. The relative thinness of the PbSe layer, on the order of 500 nm, can also partially explain the high $R_0A$ in this work.[52] These electrical characteristics are encouraging for this first demonstration of a MBE-grown PbSe-on-GaAs heterojunction, especially considering the significant dislocation density in the epilayers and the presence of the 8%-mismatched interface in the junction.

We find that PbSe-on-GaAs diodes perform well as LEDs. Room temperature, current-dependent EL spectra were taken under 5 kHz, 50% duty cycle square wave injection. In line with the $R_0A$ dependence on mesa size found previously, the small $200 \times 350$ µm mesas were found to have the brightest EL and we primarily focus on these devices. Figure 3a) plots the current dependent EL spectra from a $200 \times 350$ µm, "window contact" LED from 0.03 – 1.5 A, corresponding to current densities of 40 – 2140 A/cm$^2$ assuming uniform current spreading. The EL peak wavelength monotonically decreases with current injection from 4.0 to 3.7 µm likely due to device heating effects (note that PbSe has an inverted temperature dependence of the bandgap compared to most semiconductors).[1] The EL wavelength aligns well with the high injection PL wavelength of 3.8 µm, confirming the EL originates from recombination of electrons and holes in the PbSe layer. The inset in Figure 3a) plots the EL efficiency vs current, where the EL efficiency is the EL intensity divided by current. As the forward current is directly proportionate to the total carrier recombination rate, and the EL intensity is directly proportionate to the radiative recombination rate, the EL efficiency tracks the shape of the injection-dependent IQE.[29,47,53] At low current the EL efficiency increases rapidly up to a peak value at roughly 0.5 A (~ 700 A/cm$^2$) due to decreasing contribution from SRH recombination, and then droops at high current likely due to Auger-Meitner recombination coupled with device heating.[54] The SRH limited efficiency up to 700 A/cm$^2$ agrees well with the trap-limited ideality factor of n = 1.66 extracted from the J-V measurements, which probed a similar current density regime. Device variation was checked by measuring integrated EL intensity across 11 LEDs ($200 \times 350$ µm size) at 80 mA injection. Good uniformity in emission intensity was seen, with a less than +/- 20% variation across the devices (Figure S5).

Quantitative measurements of upper hemisphere power vs injection current, (L-I), were measured under 1 kHz, 1% duty cycle excitation. Figure 3b) shows the L-I data for a PbSe LED tested under these conditions at room temperature. The output power increases with pulsed current amplitude, reaching a value of about 400 µW at 0.8 A injection. The measured output powers for the PbSe/GaAs LEDs are greatly improved compared to the < 1 µW output powers obtained for electrically injected, homojunction PbSSe LEDs, and approach results obtained for 1W-laser optically pumped, patterned $PbSe/BaF_2$ films attached to diamond heat spreaders.[19,20] For the PbSe LEDs in this work, a peak wall plug efficiency of about 0.08% is derived. The output powers and efficiencies are already on par with some commercial LEDs in the mature III-V technology platform,[47,55] but behind the best III-V LEDs, which have efficiencies in the 0.1 – 1% range.[8,38] Still, there is room for improvement both through reducing SRH recombination to increase the IQE, and through engineering on the light extraction and injection efficiency.[56] As the PbSe epilayer is quite thin, which could lead to a reduced wall plug efficiency from poor lateral current spreading, EL intensity was measured across 200 × 350 μm LEDs with varying top contact finger spacing (Figure S6).[57] While no improvement in EL intensity was found at low injection, contact finger spacings of 12 and 18 µm led to a roughly 50% improvement in EL intensity at 571 and 1143 A/cm$^2$ injection levels compared to the window style LEDs discussed in the main text. From the device fabrication side, this suggests a clear pathway to further efficiency and output power improvement by optimizing metal contact grid separations, or by covering the entire mesa in metal and using a backside emission architecture.[47]

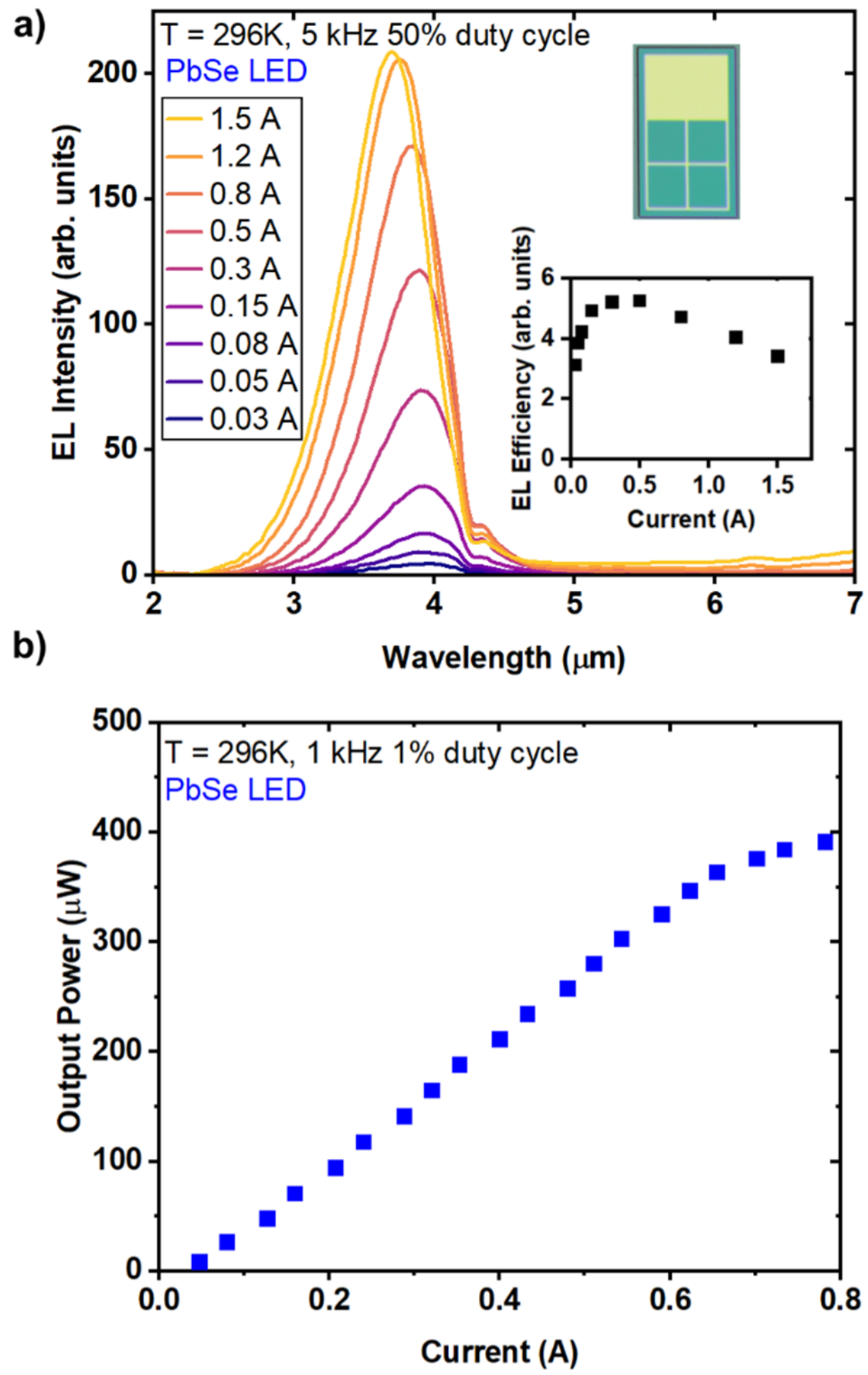

**Figure 3. a)** Current-dependent electroluminescence spectra for a 200 × 350 μm PbSe LED at room temperature. The insets show a microscopy image of the "window" style contact LED measured, and the current-dependent EL efficiency, respectively. **b)** Output power vs current for PbSe LEDs, measured under 1 kHz, 1% duty cycle injection at room temperature. Output power is seen to increase up to 400 μW at 0.8 A pulsed current injection.

We preliminarily assess if these devices degrade significantly over time. Initial aging tests using probes on unpackaged or unencapsulated devices at a high current density of 400 A/cm$^2$ with 5 kHz 50% duty cycle injection show only a 5% decline in EL intensity after 21 hours of continuous operation, (implying device lifetimes of 5 – 10 days) (Figure S7a). Subsequent low injection performance is also degraded (Figure S7b) and suggests some increase in SRH nonradiative recombination. One possible mechanism is sidewall oxidation induced degradation, as sidewalls in the current LED structures were unpassivated.[58] We aim to study the failure mechanism in future work, and compare the degradation of sidewall-passivated PbSe LEDs with the current unpassivated device design.

### 3.3. Extending LED emission to longer wavelengths with PbSnSe

The potential for wavelength tunability deeper into the mid-IR was investigated by measuring the performance characteristics of a PbSnSe LED with alloy composition tuned for emission at 5 µm. The inclusion of the wide-gap GeSe layer in-situ, though essential for protecting PbSnSe from air exposure and maintaining its IQE,[28] introduces some interesting effects. On the one hand, a potential type-I band alignment between PbSe/PbSnSe and GeSe could enhance carrier confinement, effectively forming a simple GaAs/PbSnSe/GeSe double heterostructure. On the other hand, GeSe is typically natively p-type with a low carrier concentration (around $10^{15}$–$10^{16}$ cm$^{-3}$) and poor carrier mobility,[59] whereas a LED cathode metal should ideally be deposited onto an n-type layer with high conductivity.[56] Indeed, TLM measurements of the Ti/Au contacts to the GeSe layer were found to be Schottky in nature, attributed to the poor electrical properties (Figure S3c). Figure 4a) shows the J-V characteristics for fully covered PbSnSe LED mesas. Compared to the PbSe LEDs, the leakage current density at -0.5 V is about the same order of magnitude (despite the greater TDD). However, the forward current is more severely limited by high series resistances. The inset plot of differential resistance shows that the RA of the PbSnSe devices beyond +0.5 V is between $10^{-3}$ – $10^{-2}$ Ω cm$^2$, roughly an order of magnitude higher than for the PbSe LEDs in this work. Although RTA crystallization of the amorphous GeSe cap should reduce its resistivity by orders of magnitude, the 60 nm thick crystalline GeSe layer is still relatively insulating and leads to a large series resistance penalty.[60] The zero-bias $R_0A$ for the PbSnSe devices is on the order of about 30–40 Ω cm$^2$. Although the epitaxial layer stack is about half the thickness as for the PbSe devices, the higher $R_0A$ could also be a sign of a double heterojunction effect with the wider gap GaAs and GeSe layers sandwiching the PbSe and PbSnSe layers.

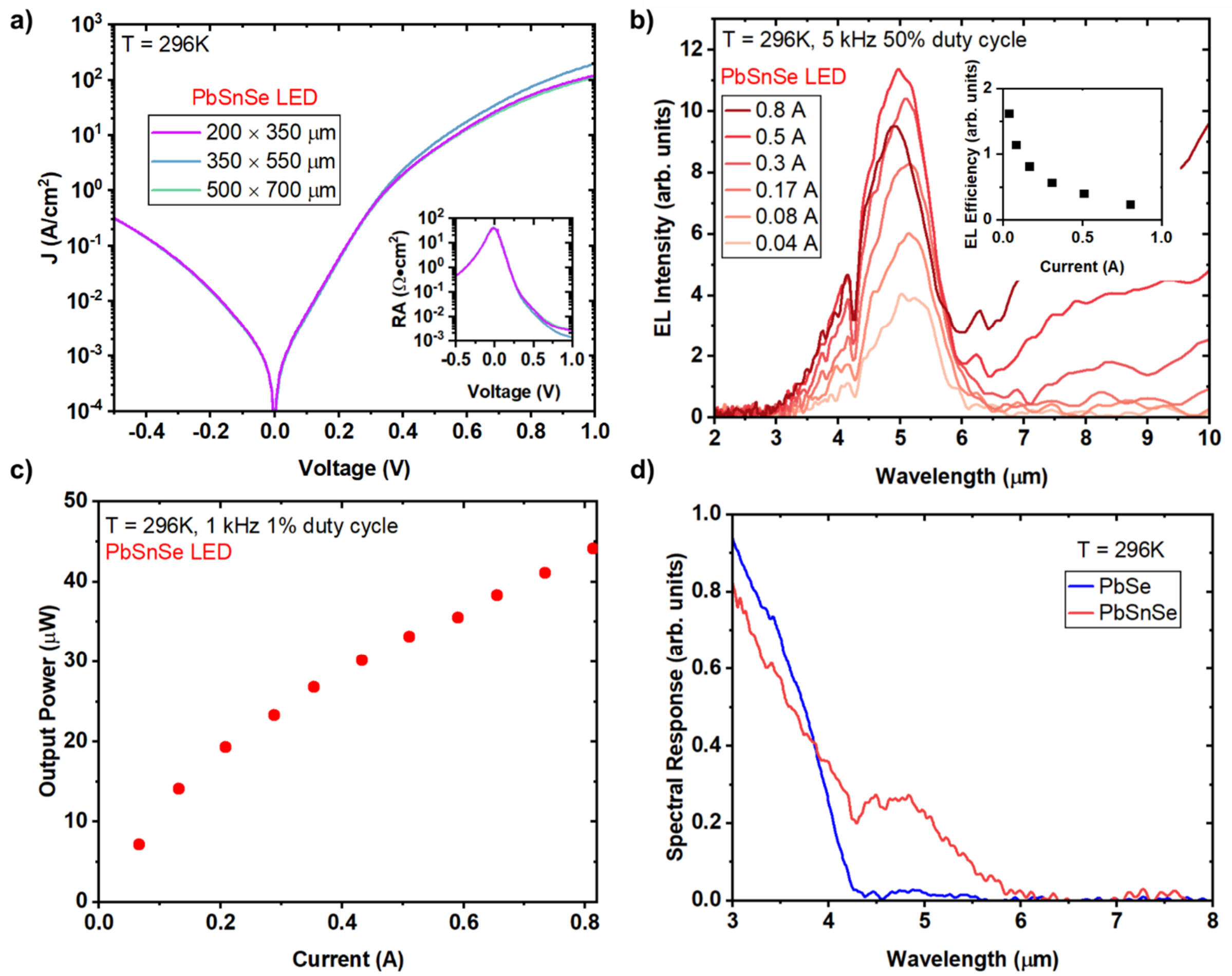

**Figure 4. a)** Room temperature J-V curves for PbSnSe LEDs with varying mesa sizes, with the applied voltage swept from -0.5 – 1 V. The forward current density in the PbSnSe LEDs at 0.5 V is roughly 10× lower than for the PbSe/GaAs devices, due to a higher series resistance from the low conductivity GeSe passivation layer. The inset shows the differential resistance area products for the diodes. **b)** Current-dependent electroluminescence for the 350 × 550 μm PbSnSe LED at room temperature. The inset shows the rapidly decreasing EL efficiency with current. **c)** Light output power vs current for PbSnSe LED, measured under 1 kHz, 1% duty cycle current injection to reduce device heating effects. LED output power increases from 10 – 45 μW as the pulsed current is increased from 0.1 – 0.8 A. **d)** Zero-bias spectral response from the PbSe and PbSnSe devices. Photovoltaic detection is seen at wavelengths shorter than 4 μm for the PbSe LED, while the photoresponse is extended past 5 μm for the 7% Sn, PbSnSe LED.

Figure 4b) plots the current-dependent EL spectra for a PbSnSe LED under 5 kHz, 50% duty cycle testing at room temperature. The EL peak wavelength of 5 μm matches with the PL data for the PbSnSe epilayer, confirming the wavelength extension with Sn alloying for electrically injected emission. To our knowledge, this is the first demonstration of room temperature EL from a PbSnSe LED. Under the 50% duty-cycle testing conditions, the high resistance GeSe layer leads to a severe device heating effect that greatly decreases emission efficiency at higher currents. The heating is visible in the EL spectra as the increasing signal beyond 7 μm. While the EL efficiency of the PbSnSe LED is comparable to the PbSe LED at low current (Figure 4b) inset), the PbSnSe LED is > 10× dimmer at higher currents. The poor performance is not intrinsic to the PbSnSe epilayer, however, as the PL intensity at high optical excitation is only about a factor of 2× worse than the PbSe sample and points to room for improvement in the device design (Figure 1d).

Under 1% duty cycle testing, the device heating during excitation is greatly reduced. Figure 4c) plots the L-I data for the PbSnSe LED, which reaches an output power of 45 μW at 0.8 A injection, well beyond the nW level output powers in previous PbSnSe bulk homojunction LEDs.[17] While the wall plug efficiency still monotonically decreases with injection, a peak value at low current of 0.015% is estimated under these testing conditions. The wall plug efficiency is comparable to recent results on 5 μm emitting CdSe and 6 μm emitting antenna-enhanced HgTe colloidal quantum dot LEDs, where wall plug efficiencies of 0.013% and 0.036% were extracted, respectively.[14,15] However, the PbSnSe LED is still about an order of magnitude less efficient than the best III-V LEDs at 5 μm, which have wall plug efficiencies on the order of about 0.2% at room temperature.[61] Nevertheless, these early results demonstrate the potential of PbSnSe for emission at challenging wavelengths beyond 4 μm. If suitable wide-gap, n-type, and oxidation resistant replacement layers to GeSe can be found, the wall plug efficiency could be greatly improved by reducing the parasitic series resistance. As with the PbSe LEDs, further improvements in IQE and extraction efficiency should also be possible. Given the success of PbSe colloidal quantum dots as near-IR emitters, together with the demonstrated mid-IR electroluminescence of HgTe and CdSe quantum dot LEDs,[62] colloidal quantum dots based on PbSnSe ternary alloys may represent an additional pathway toward longer wavelength emission. Indeed, absorption edge tunability extending into the mid-IR has already been demonstrated in related PbSnTe colloidal quantum dot systems.[63]

Finally, in addition to light emission applications, the fabricated PbSe and PbSnSe LEDs function as photodetectors when a reverse bias is applied to the LED. Figure 4d) plots the wavelength dependent photoresponse of the PbSe and PbSnSe LED, respectively. Light detection is observed at zero bias at wavelengths shorter than 4 µm for the PbSe LED, with photoresponse extending to 6 µm for the Sn-alloyed PbSnSe LED. Although the epitaxial layer thicknesses in this work are much thinner than needed for full absorption,[64] the initial photoresponse demonstrates the potential for harnessing dissimilar semiconductors in hybrid IV-VI/III-V heterojunctions for both emission and detection applications across the mid-IR spectrum.

## 4. DISCUSSION

Table 1 benchmarks the PbSe and PbSnSe/GaAs LEDs in this work against incumbent III–V mid-IR LEDs and recent results from other emerging material systems.

**Table 1.** Comparison of IV-VI/III-V LED performance characteristics with recent literature results for mid-IR LEDs at room temperature. Emission wavelength $\lambda_{peak}$, maximum reported output power, and the maximum wall plug efficiency (WPE).

| Device architecture | $\lambda_{peak}$ (µm) | Max. reported output power (µW) | Max. WPE (%) | Reference |
|---|---|---|---|---|
| III-V ICLED | 3.7 | 5100 | 0.7 | 8 |
| bP | 3.7 | 8 | 1.78 | 13 |
| **PbSe/GaAs** | **3.8** | **400** | **0.08** | **this work** |
| III-V C-SLED | 4.0 | 6800 | 0.3 | 65 |
| Hamamatsu (III-V) | 4.3 | 3100 | 0.6 | 38 |
| **GeSe/PbSnSe/GaAs** | **5.0** | **45** | **0.015** | **this work** |
| CdSe Quantum Dot | 5.0 | 195 | 0.016 | 14 |
| Nanoplus(III-V ICLED) | 5.1 | 820 | 0.2 | 61 |
| III-V ICLED | 5.4 | 600 | 1.2 | 66 |
| HgTe Quantum Dot | 6.0 | 200 | 0.036 | 15 |

The PbSe and PbSnSe/GaAs hybrid LEDs perform competitively with emerging black phosphorus and II-VI colloidal quantum dot LEDs and, notably, are already within a factor of 10× of the best III-V LEDs in terms of room temperature power and efficiency at similar mid-IR wavelengths. Coupled with sub-400 °C growth and fabrication temperatures, direct heteroepitaxial growth and integration of IV-VI emitters on lattice-mismatched III-V and Si/Ge mid-IR photonic platforms is a compelling possibility.[67,68] Nevertheless, more work is needed to further optimize materials growth to reduce SRH such that devices at high injection are truly limited by the low Auger-Meitner coefficients of IV-VI materials. One potential pathway is to grow (111) and/or (110) oriented IV-VI device epilayers. As the Schmid factor for the primary <110>{100} slip system becomes nonzero for (111) and (110) oriented films, unlike for (001) films, thermal cyclic annealing can lead to significant dislocation motion and annihilation, and subsequent orders of magnitude reductions in TDD from the currently high TDD of $10^9$ $cm^{-2}$.[69] Recombination at dislocations and point defects might be slower than that in III-V semiconductors, but these materials are certainly not insensitive to defects. Reducing nonradiative recombination at defects may also hold the key to improved device reliability.

In addition to reducing nonradiative defects, further improvements in wall plug efficiency are expected through device structure optimization and optical engineering. Increasing the PbSe active layer thickness beyond 475 nm toward 1 – 2 µm, as is typical for III-V bulk heterojunction LEDs, should improve current spreading and shift peak IQE to higher current densities by delaying the onset of Auger-Meitner recombination induced efficiency droop, although this must be balanced against increased photon reabsorption losses in thicker active regions.[47,56,57] However, increasing the active layer thickness should certainly be beneficial for the PbSnSe LED, which presently employs a much thinner 150 nm active region, as Auger-Meitner recombination is expected to be more severe than in the wider bandgap PbSe LED.

From an optical engineering standpoint, light extraction is severely limited by the high refractive indices of the PbSe film (n = 5) and GaAs substrate (n = 3.3). For planar thin films, top-side emission is governed by total internal reflection, with extraction efficiency scaling as approximately $1/4n^2$,[56] especially problematic for high index semiconductors like PbSe. Implementing a backside emission architecture, combined with metalens-based outcoupling strategies, has the potential to substantially enhance light extraction from the active region of the LED.[47,68] Additionally, nanophotonic patterning of the LEDs could exploit the ultra-high refractive indices of PbSe and PbSnSe to further boost optical performance.[71,72]

Considering longer wavelength LEDs, the output power of the PbSnSe LEDs is strongly limited by parasitic series resistance from the insulating GeSe barrier layer, rather than the low IQE of the active region. Therefore, replacing GeSe with a suitable wide-gap, n-type, and oxidation resistant barrier layer is essential to reduce the series resistance and thereby improve both wall plug efficiency and output power. One promising candidate is $Bi_2S_3$, a wide-gap, intrinsically n-type semiconductor that has already shown potential in $PbS/Bi_2S_3$ heterojunction solar cells.[73,74] Additionally, the emission wavelength of PbSnSe LEDs can also be extended much deeper into the mid-IR by further increasing the Sn composition of the active layer. In previous work on PbSnSe heterostructures, high injection PL intensity was observed to only decrease by a factor of 3× with emission wavelength redshift from 4.3 to 7 µm, corresponding to a shift in the active layer alloy composition from 5 – 17% Sn.[28] These relatively modest losses, along with the strong PL intensity of the 5 µm emitting PbSnSe epilayer relative to the PbSe epilayer in this work, suggest that output powers beyond 10 µW should be readily achievable beyond 5 µm solely by increasing the Sn composition. As with the PbSe LEDs, further improvements in active region IQE and light extraction efficiency should also be possible for PbSnSe devices. Together, these pathways highlight the viability of hybrid IV–VI/III–V heterojunction LEDs as a scalable, wavelength-flexible mid-IR light source platform that can bridge the performance and cost gap between incumbent laser and thermal emitter technologies.

One clear disadvantage of PbSe/PbSnSe-based LEDs relative to III-V LEDs and thermal emitters is the high toxicity of lead. In practice, however, this concern can be mitigated through careful device preparation, handling, and encapsulation strategies that are already widely employed in commercially available PbS and PbSe infrared detectors.

## 5. CONCLUSION

A hybrid IV-VI/III-V LED platform with PbSe and PbSnSe heterostructures on GaAs was investigated at room temperature. Despite threading dislocation densities on the order of $10^9$ $cm^{-2}$, these LEDs emit at 3.8 and 5 µm, with output powers under 1% duty cycle testing up to 400 and 45 µW for a PbSe and 7% Sn PbSnSe LED, respectively. The heterojunction with the wide bandgap GaAs substrate leads to greatly improved device performance compared to previous reports on homojunction PbSe diodes. Peak wall plug efficiencies of 0.08% and 0.015% are determined for the PbSe and PbSnSe LEDs, respectively, which are competitive with many existing commercial LEDs at the same wavelength. Importantly, there is still room for improvement in the IQE, light extraction efficiency, and current spreading in both emitters. Currently, the performance of PbSnSe

LEDs is also limited by the requirement to passivate the surface with relatively insulating GeSe, which leads to a large series resistance penalty, and thus increased device heating. More suitable wide-gap passivation layers will need to be explored in future work. Overall, these results demonstrate a step change in the performance of electrically injected IV-VI LEDs and establish hybrid IV-VI/III-V heterojunction LEDs as a viable path towards low cost, compact, efficient, and tunable mid-IR emitters for room temperature sensing, imaging, and spectroscopy.

## SUPPORTING INFORMATION

Details about RHEED during epilayer growth, x-ray rocking curves before and after annealing, Sb doping and metal contacts, device uniformity, current spreading, and preliminary device aging.

## ACKNOWLEDGEMENTS

We gratefully acknowledge support via the NSF CAREER award under Grant No. DMR-2036520 and the UC Santa Barbara NSF Quantum Foundry funded via the Q-AMASE-I program under Award No. DMR-1906325. Part of this work was performed at the Stanford Nano Shared Facilities (SNSF) and Stanford Nanofabrication Facilities (SNF), supported by the National Science Foundation under Award No. ECCS-2026822. J. E. M gratefully acknowledges support from the Tomkat Center for Sustainable Energy's Tomkat Center Graduate Fellow for Translational Research Fellowship. L.N. gratefully acknowledges support from the Geballe Laboratory for Advanced Materials Postdoctoral Research Fellowship. L.N. and B.R. also acknowledge support from Dr. Joshua Lentz at the Air Force Research Laboratory and the NSF I/UCRC on Multi-functional Integrated System Technology (MIST) Center, Award No. 1939050.

## CONFLICT OF INTEREST

The authors declare no conflict of interest.

# Supplementary Information

## Mid-infrared LEDs based on lattice-mismatched hybrid IV-VI/III-V heterojunctions

Jarod E. Meyer[1], Biridiana Rodriguez[2], Leland Nordin[2,3*], and Kunal Mukherjee[1†]

[1]Department of Materials Science and Engineering, Stanford University, Stanford, CA 94305, USA

[2]CREOL, The College of Optics and Photonics, University of Central Florida, Orlando, FL 32816, USA

[3]Department of Materials Science and Engineering, University of Central Florida, Orlando, FL 32816, USA

### 1. RHEED of PbSe and PbSnSe LED epilayer stacks

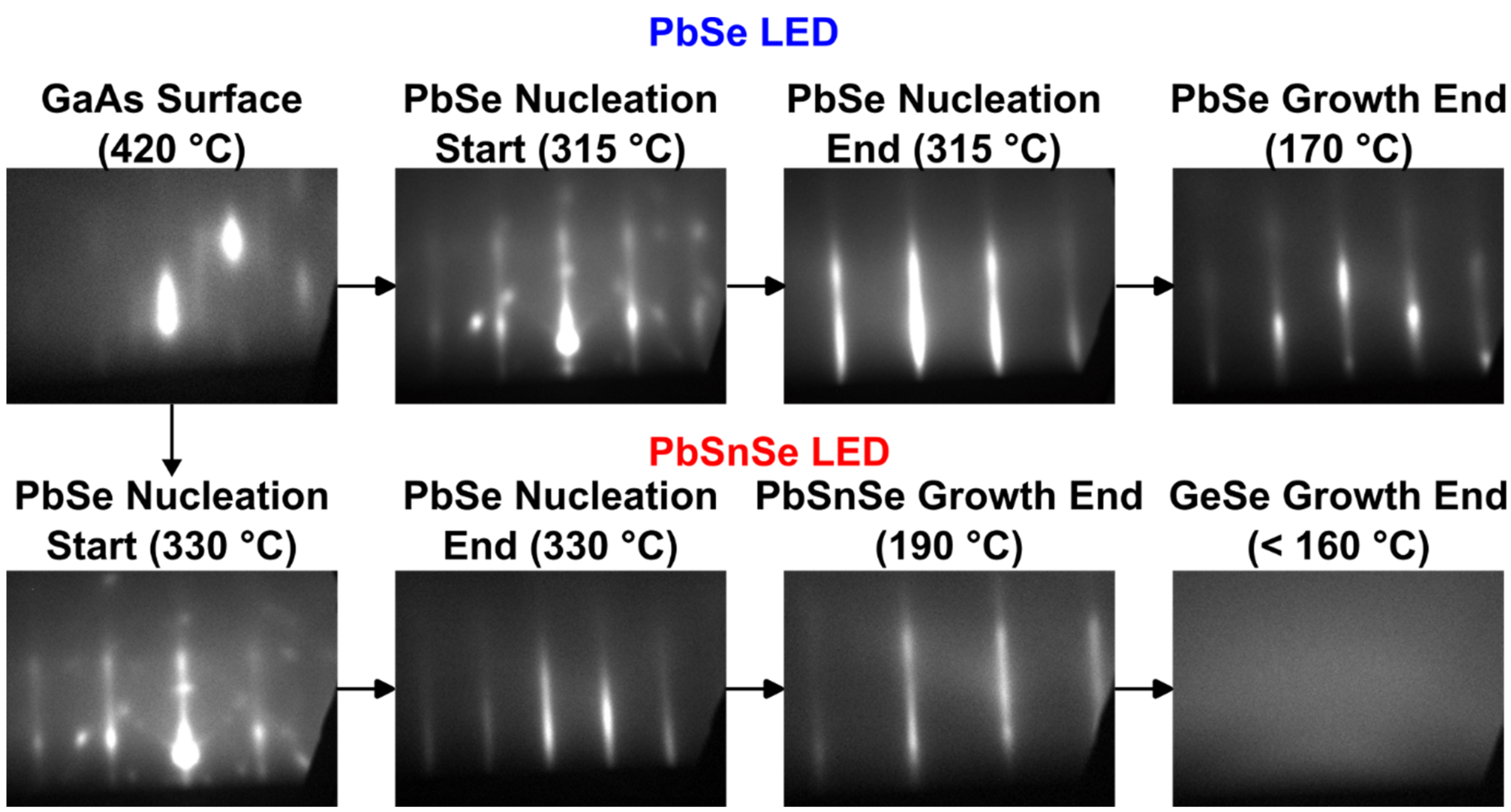

**Figure S1.** RHEED images of the epilayer surfaces during various growth steps for the PbSe and PbSnSe epilayer stacks. The RHEED pattern of the GaAs surface at 420 °C after dosing with PbSe looks identical between the two epilayers. PbSe nucleation at 315 – 330 °C shows initially spotty RHEED, but eventually a streaky 1 × 1 pattern emerges after roughly 25 nm of deposition. Some roughening is observed in RHEED for PbSe/PbSnSe growth below 200 °C, though the growth remains single orientation. Finally, for the PbSnSe LED epilayer, RHEED becomes hazy and diffuse due to deposition of the amorphous GeSe capping layer.

* leland.nordin@ucf.edu
† kunalm@stanford.edu

## 2. XRD rocking curves before and after annealing

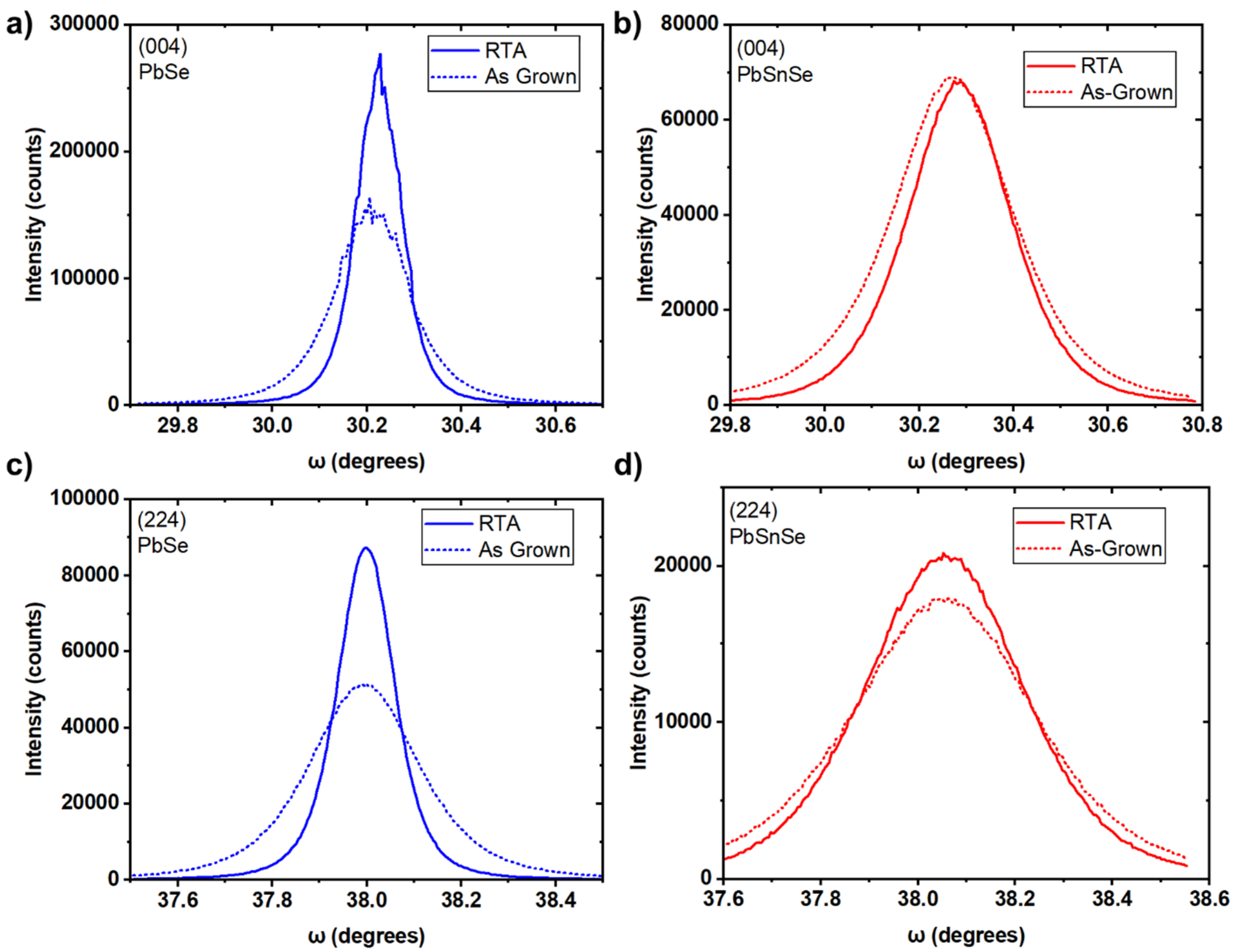

**Figure S2. a - b)** Symmetric, (004) rocking curves of the PbSe and PbSnSe LED epilayers, respectively, before and after annealing. **c - d)**, Skew symmetric, (224) rocking curves.

Figure S2 a – b) plots the symmetric (004) rocking curves of the PbSe and PbSnSe LED epilayers before and after SiO2-capped rapid thermal annealing for 300 s at 375 °C. The out-of-plane crystallinity of the PbSe epilayer is greatly improved with annealing, as the (004) full-width-at-half-maximum (FWHM), decreases from 670 to 410 arcseconds. Meanwhile, for the PbSnSe epilayer, a slighter improvement from 1120 to 950 arcseconds is observed. Similar trends are observed in the skew-symmetric (224) rocking curves, (Figures S1c-d), which also capture the in-plane crystallinity. For the PbSe epilayer, the (224) FWHM decreased from 1010 to 510 arcseconds, while for the PbSnSe epilayer a decrease from 1700 to 1430 arcseconds was observed.

## 3. Sb doping and metal contacts to device epilayers

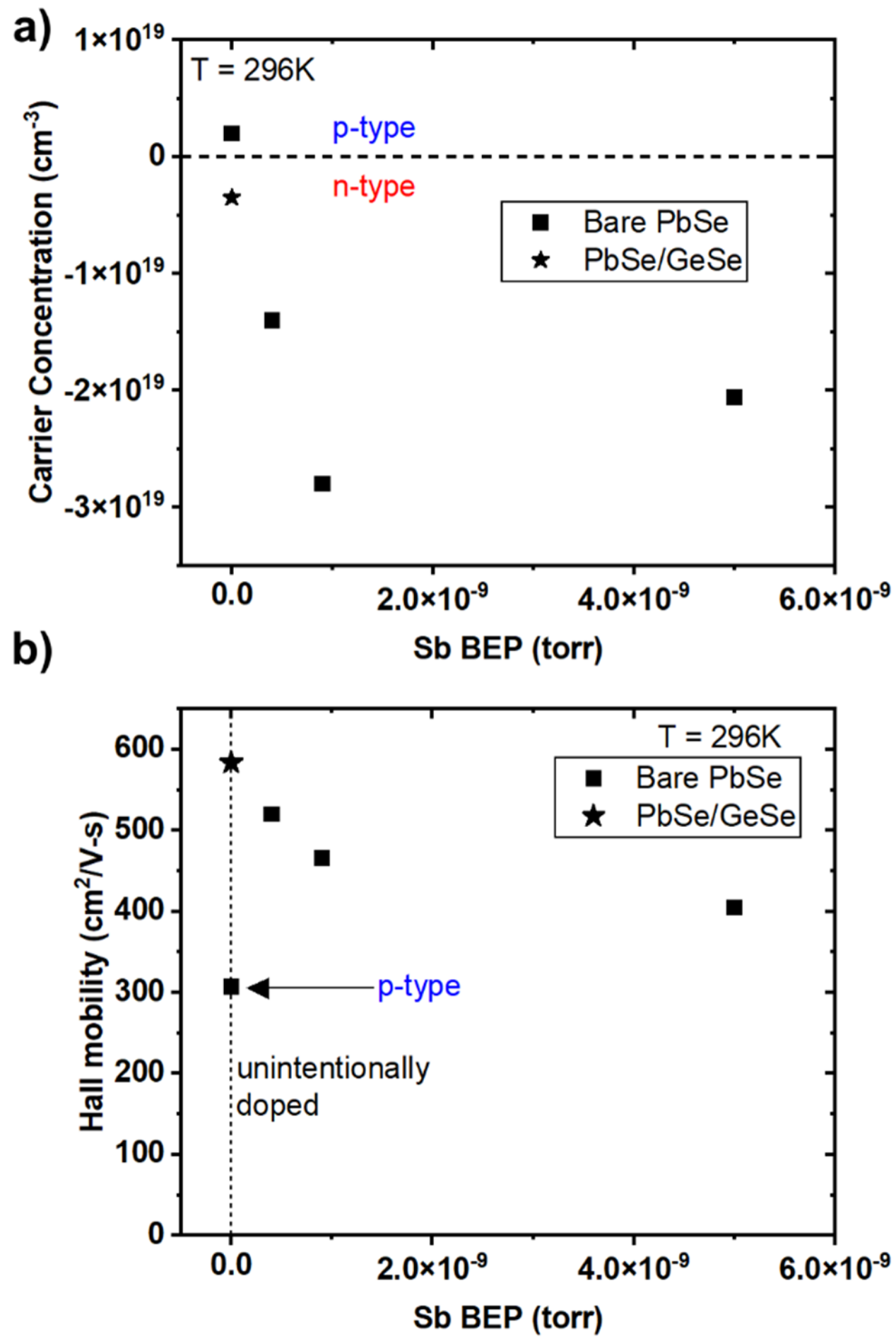

**Figure S3. a)** Hall carrier concentration for bare PbSe thin films with increasing Sb BEP. The star symbol shows the hall data for an unintentionally doped PbSe thin film that was in-situ capped with amorphous GeSe, which possess an n-type carrier concentration in the mid $10^{18}$ $cm^{-3}$ range. **b)** Hall mobility for the same set of films in a). n-type films show a higher carrier mobility than for the bare, p-type PbSe film, though the mobility monotonically decreases with increasing Sb BEP.

n-type doping with Sb was investigated by growing PbSe thin films on semi-insulating GaAs substrates and introducing an additional Sb flux from a valved cracker cell. Figure S3) shows the Hall carrier concentration (Figure S3a), and Hall mobility (Figure S3b), for PbSe layers with an increasing Sb beam-equivalent-pressure (BEP) up to $5 \times 10^{-9}$ torr.

Similar to our previous work, bare thin films of PbSe show a p-type carrier concentration in the low $10^{18}$ $cm^{-3}$ range, due to oxygen exposure, while films in-situ capped with GeSe show n-type carrier densities in the $10^{18}$ $cm^{-3}$ range.[1] With increasing Sb BEP, the carrier type of bare PbSe is switched n-type with densities up to $3 \times 10^{19}$ cm$^{-3}$ for a Sb BEP of $10^{-9}$ torr. Further increase in Sb BEP does not lead to a larger doping background, however, likely due to compensation effects.[2] The hall mobility for the n-type films, achieved through GeSe surface passivation or Sb doping, are found to be higher than for the bare, unintentionally doped p-type PbSe film. Increasing Sb doping leads to a monotonic reduction in mobility, however.

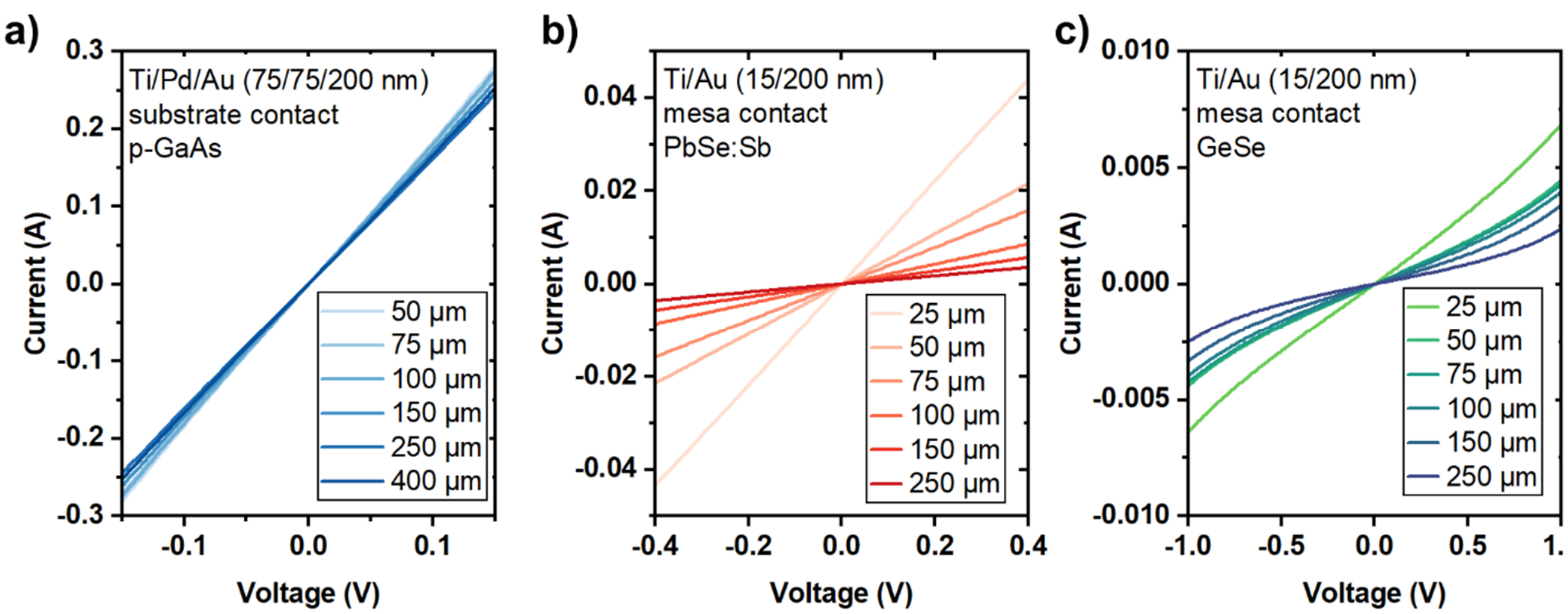

**Figure S4. a)** I-V curves for Ti/Pd/Au metal contacts to the p-GaAs substrate. **b)** I-V curves for Ti/Au contacts to PbSe:Sb in the PbSe LED devices. **c)** I-V curves for Ti/Au contacts to the GeSe top layer in the PbSnSe LED devices.

The metal-semiconductor contact characteristics are investigated via I-V measurements on $220 \times 150$ µm rectangular metal contacts with increasing separation, i.e. the so-called transmission line method. Figures S4a – c) plot the I-V curves for Ti/Pd/Au metal contacts to the p-GaAs substrate, and for Ti/Au metal contacts to PbSe:Sb and GeSe, respectively.

In Figures S4a) and S4b), Ti/Pd/Au and Ti/Au form ohmic contacts with linear I-V curves to p-type GaAs and Sb doped, n-type PbSe, respectively. For the relatively insulating GeSe layer, however, Ti/Au results in Schottky contacts with high resistivity, as shown in Figure S4c).

## 4. Device uniformity

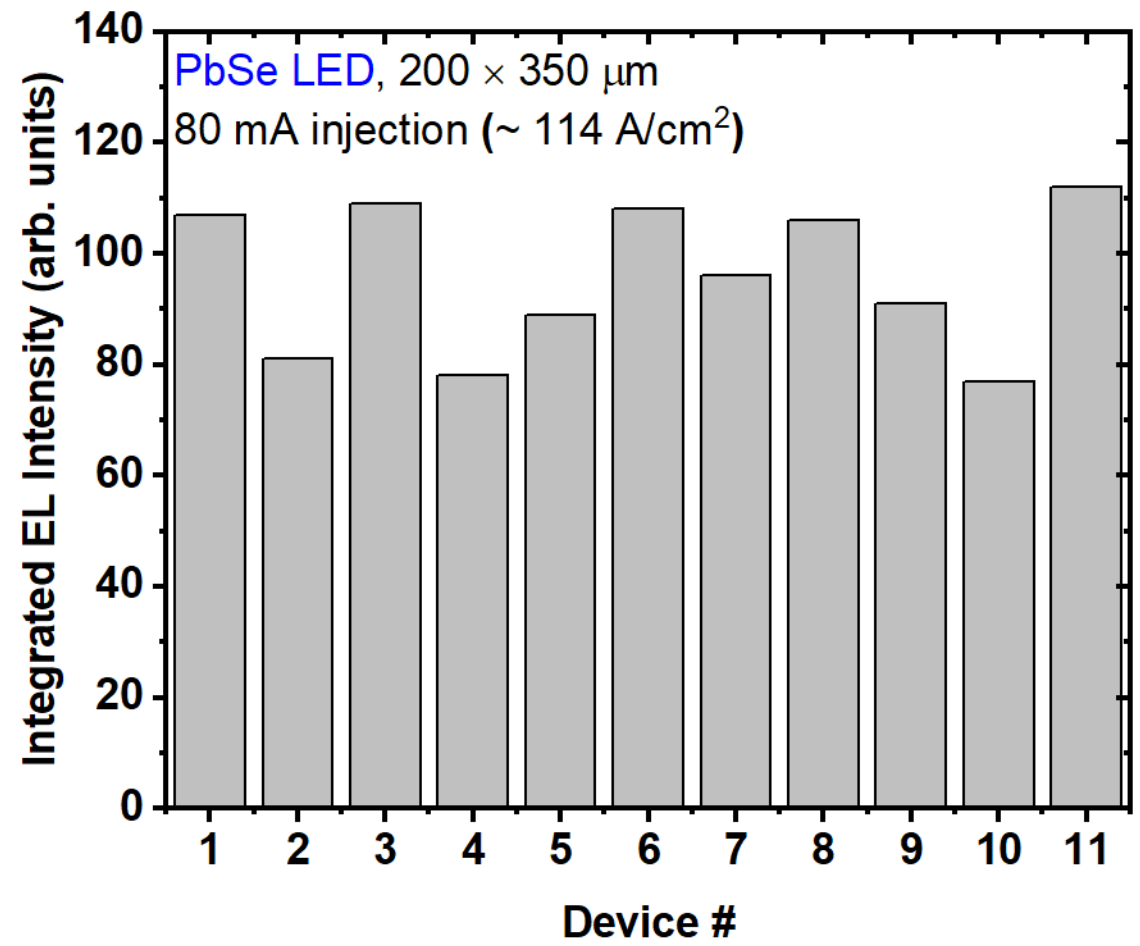

**Figure S5.** Integrated electroluminescence intensity across 11 PbSe LEDs at a low injection current of 80 mA.

Device uniformity across a 1 cm × 1 cm chip was evaluated by measuring electroluminescence (EL) from 11 200 × 350 µm PbSe LED mesas. A low current of 80 mA was chosen such that the luminescence efficiency would be limited by nonradiative defect recombination, which could vary across a wafer, rather than intrinsic mechanisms. Figure S5) shows the electroluminescence intensity across the 11 devices. The EL intensity is seen to vary by only about 20% across the 11 PbSe LEDs, suggesting good device uniformity for this early-stage work.

## 5. Impact of current spreading

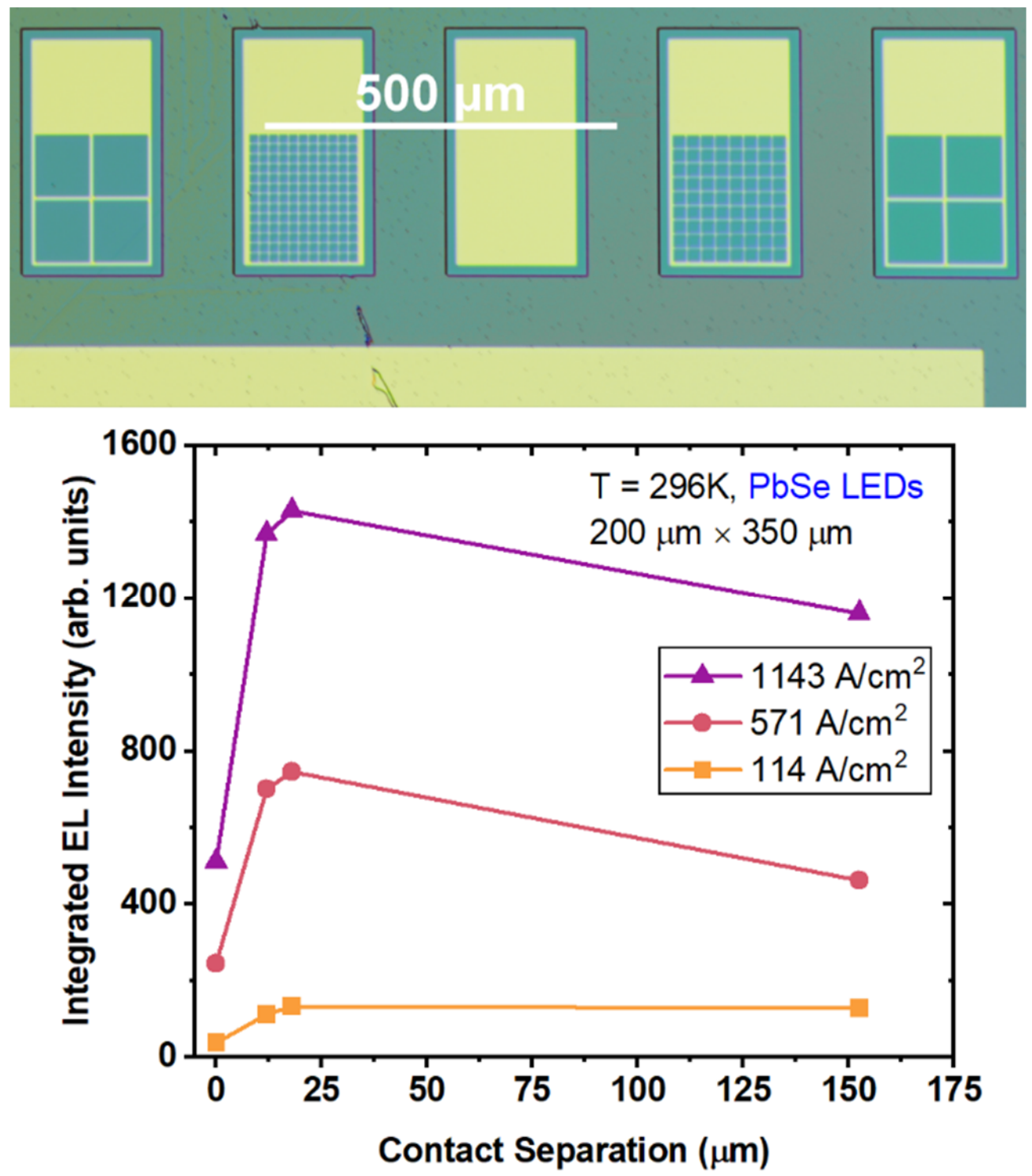

**Figure S6.** EL intensity vs metal contact finger separation for PbSe LEDs at 3 current injection levels. The top microscopy image shows the 4 contact geometries investigated: an open "window-style" mesa contact, metal grids of 2 µm lines separated by 12 and 18 µm spaces, and mesas completely covered with metal except for a 15 µm perimeter. The lines are solely for guiding the eye between the different injection levels

The impact of current crowding at metal contacts on EL intensity was investigated via current-dependent EL on PbSe LED mesas with varying top-contact geometries, as shown in Figure S6. Compared to the open window-style contacts investigated in the main text, the EL intensity of mesas patterned with metal lines separated by 12 µm and 18 µm spaces, and mesas fully covered with metal except for the 15 µm free perimeter, ("0 µm separation"), were also measured . At low current densities of roughly 114 A/cm$^2$, no improvement in EL intensity was found for the smaller metal contact spacings, suggesting current crowding does not significantly impact the luminescence in this regime. At higher current densities however, the current crowding effect becomes more significant.[3,4] At 571 and 1143 A/cm$^2$, a roughly 50% improvement in EL intensity is seen for the 12 and 18 µm contact spacings, despite the mesas having less open area for light emission. More work is needed to determine the optimal contact finger spacing, which is likely between 18 and 150 µm for these LEDs.

## 6. Preliminary Device Aging

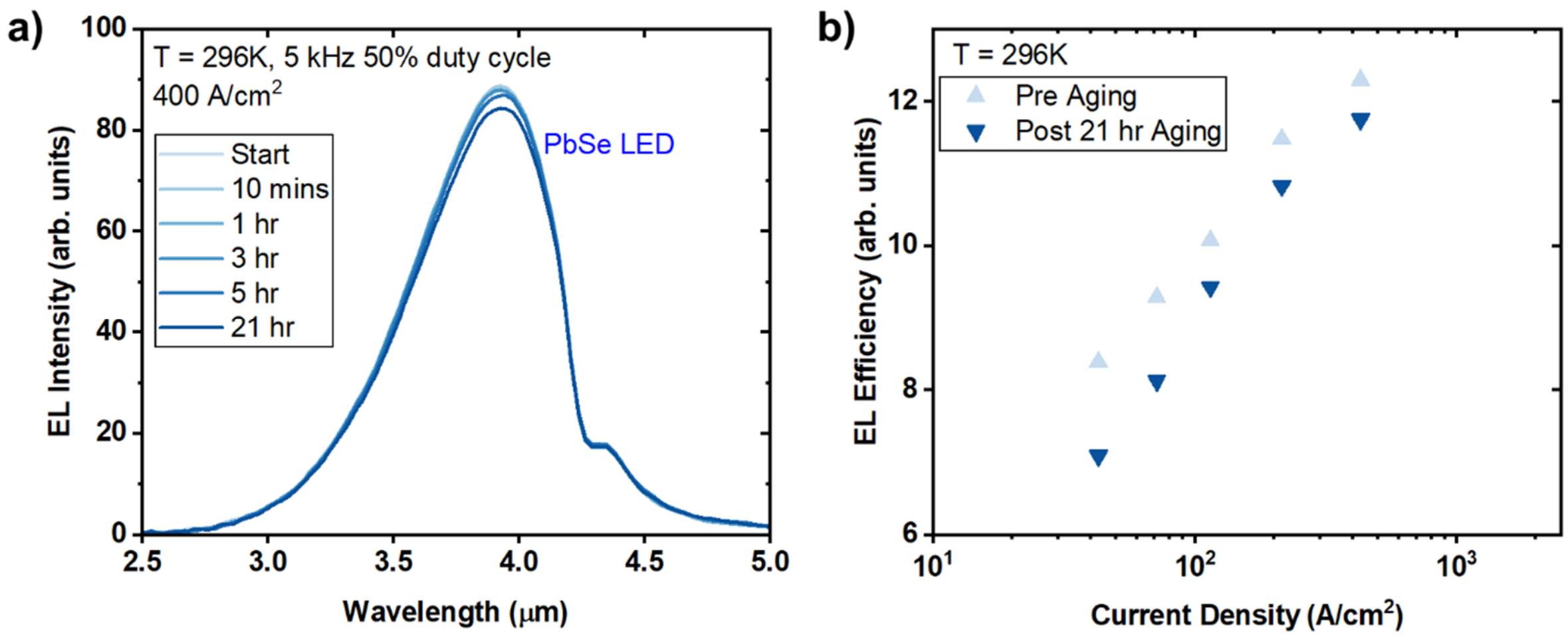

**Figure S7. a)** Electroluminescence spectra of PbSe LED over 21 hours of continuous operation at 400 A/cm$^2$, using 5 kHz, 50% duty cycle current pulsing. **b)** EL Efficiency before and after the 21 hours of aging. The degradation in EL Efficiency, (and thus EL intensity), is significantly worse at lower currents, suggesting a rise in the strength or number of nonradiative SRH recombination centers.

The reliability of a PbSe LED was preliminarily tested under 5 kHz, 50% duty cycle operation at a high current density of 400 A/cm$^2$. The LED was ran overnight for 21 hours, with EL spectra taken at various intervals. Figure S7a) shows the time dependent EL spectra of the PbSe LED during aging.

After 21 hours of aging, the EL intensity of the PbSe LED is roughly 95% of its original value, implying a device lifetime on the order of 5 – 10 days. EL Efficiency (EL Intensity divided by current), was measured before and after aging, as shown in Figure S7b). The EL Efficiency at low current is more significantly degraded than at higher injection, suggesting an increase in SRH recombination after 21 hours of aging is responsible for the degradation in EL intensity. The exact aging mechanism, however, requires further investigation.